\documentclass{PoS}

\title{Fitting the Fermi-LAT GeV excess: on the importance of the propagation of electrons from dark matter}

\ShortTitle{Fitting the Fermi-LAT GeV excess: on the importance of the propagation of electrons from dark matter}

\author{\speaker{Thomas Lacroix}\\
        UPMC-CNRS, UMR7095, Institut d'Astrophysique de Paris, 98 bis boulevard Arago, 75014 Paris, France\\
        E-mail: \email{lacroix@iap.fr}}

\abstract{An excess of gamma rays at GeV energies has been detected in the Fermi-LAT data. This signal comes from a narrow region around the Galactic Center and has been interpreted as possible evidence for light (30 GeV) dark matter particles. Focussing on the prompt gamma-ray emission, previous works found that the best fit to the data corresponds to annihilations proceeding into $b$ quarks, with a dark matter profile going as $r^{-1.2}$. We show that this is not the only possible annihilation set-up. More specifically, we show how including the contributions to the gamma-ray spectrum from inverse Compton scattering and bremsstrahlung from electrons produced in dark matter annihilations, and undergoing diffusion through the Galactic magnetic field, significantly affects the spectrum for leptonic final states. This drastically changes the interpretation of the excess in terms of dark matter.}

\FullConference{Frontiers of Fundamental Physics 14\\
		 15-18 July 2014\\
		 Aix Marseille University (AMU) Saint-Charles Campus, Marseille, France}

\begin{document}

\section{Introduction: the Galactic Center gamma-ray excess}

An excess of gamma rays from the Galactic Center (GC) has been detected in the Fermi-LAT data between roughly 0.1 and 10 GeV \cite{Fermi_collaboration,Hooper_Linden_excess,Gordon_Macias_excess,Abazajian_GeV_excess,Daylan_GeV_excess}. This excess has a spatial extension smaller than $10^\circ \times 10^\circ$, and is spherically symmetric. It was obtained by subtracting to the data known sources and a template for the background diffuse emission provided by the Fermi Collaboration. Although this background modelling procedure has been debated, the picture that has emerged seems to be robust. There is a variety of astrophysical explanations for the excess, but an interpretation in terms of dark matter (DM) is nevertheless possible.

\begin{figure}[!ht]
\centering
\includegraphics[scale=0.37]{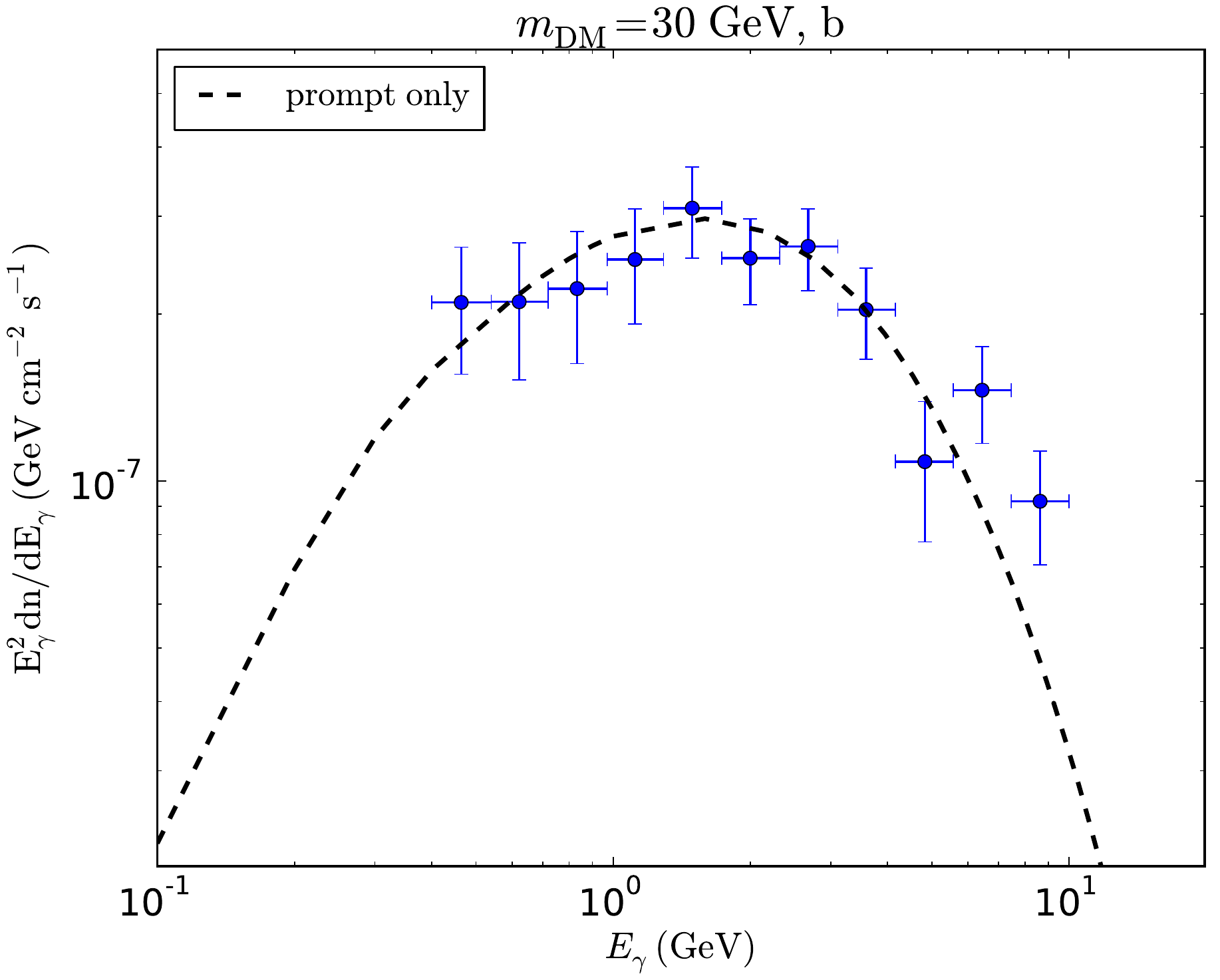}
\includegraphics[scale=0.37]{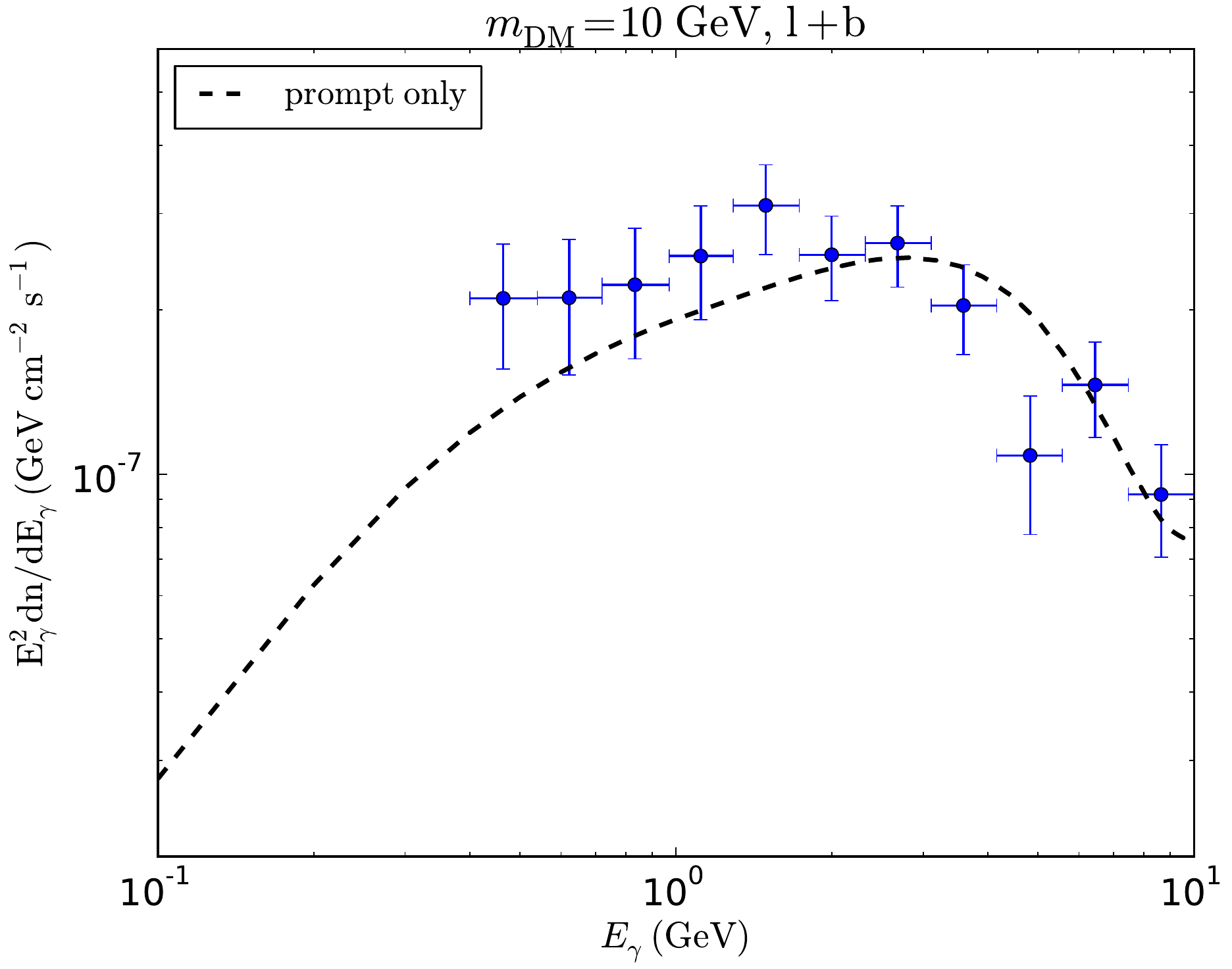}
\includegraphics[scale=0.37]{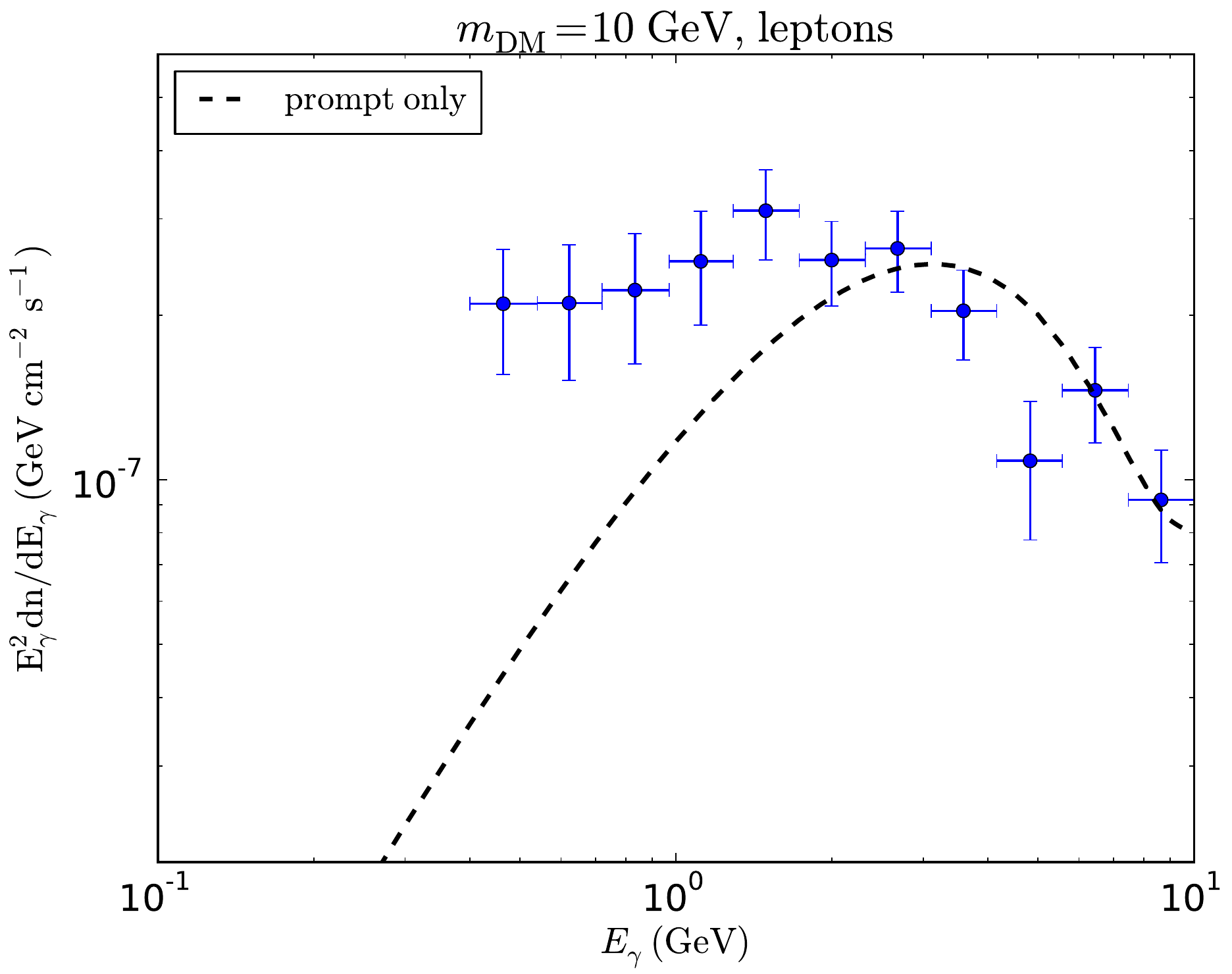}
\caption{Best fit to the spectrum of the residual extended emission in the $7^{\circ} \times 7^{\circ}$ region around the GC, for 30 GeV DM annihilating into 100\% $b$ quarks (top left panel), and 10 GeV DM annihilating into 90\% leptons and 10\% $b$ quarks (top right panel), and 100\% leptons (bottom panel). The best-fit cross section is $\sim 2 \times 10^{-26}\ \rm cm^{3}\ s^{-1}$. The data points are taken from \cite{Gordon_Macias_excess}.}
\label{prompt_fits}
\end{figure}

The best fit to the excess quoted in the literature has been obtained for 30 GeV DM annihilating into $b\bar{b}$ (Fig.~\ref{prompt_fits}, top left panel), with a cross section of $2 \times 10^{-26}\ \rm cm^{3}\ s^{-1}$. The data also point to a density profile $\rho \propto r^{-1.2}$. A mixture of 90\% leptons and 10\% $b$ quarks gives a relatively good fit (Fig.~\ref{prompt_fits}, top right panel), while the fit is bad for a final state containing 100\% leptons (Fig.~\ref{prompt_fits}, bottom panel). Here the term ``leptons'' refers to democratic annihilation into leptons, i.e.~a combination of the $e^{+}e^{-}$, $\mu^{+}\mu^{-}$, $\tau^{+}\tau^{-}$ final states, with 1/3 of the annihilations into each of these channels. 

Consequently, final states of DM annihilation containing only leptons have not been considered viable when interpreting the GeV excess in terms of DM. However, these conclusions were obtained taking into account only the prompt gamma-ray emission, namely the final-state radiation (FSR) single-photon emission, and the immediate hadronization and decay of the DM annihilation products into photons. Nevertheless, electrons and positrons are also by-products of DM annihilations, and they produce gamma rays via inverse Compton (IC) scattering off photons of the interstellar radiation field and bremsstrahlung. It had been argued by the authors of Refs.~\cite{Fermi_IC,brems_Cirelli} that the contributions of these secondary emissions to the gamma-ray spectrum should not be neglected and should lead to corrections. On top of that, diffusion of electrons and positrons through the Galactic magnetic field must be taken into account when modelling these emissions. Here we show that these contributions do not just give corrections but they totally change the interpretation of the excess in terms of DM.

\section{Diffusion of electrons and positrons from DM}

In order to compute the gamma-ray spectrum from DM-induced electrons and positrons, we solve the diffusion-loss equation of cosmic rays. Assuming a steady state, the equation reads \cite{Itilde_Timur}
\begin{equation}
K \nabla^{2}\psi + \frac{\partial}{\partial E}(b_{\mathrm{tot}} \psi) + q = 0,
\end{equation}
where $ \psi(\vec{x},E) $ is the cosmic-ray spectrum after propagation, $ K $ is the diffusion coefficient: $ K(E) = K_{0} \left( E/E_{0} \right) ^{\delta} $ with $ E_{0} = 1 \ \rm GeV $, $ b_{\mathrm{tot}}(E) $ is the total energy loss rate (IC, synchrotron, bremsstrahlung) and $ q(\vec{x},E) $ is the source term for DM annihilations, proportional to $ \rho^{2} $. To solve the equation, we use the semi-analytic method described in Ref.~\cite{Itilde_Timur}. In this approach, the spectrum of electrons and positrons after diffusion is given by 
\begin{equation}
\psi (\vec{x},E) = \frac{\kappa}{b_{\mathrm{tot}}(E)} \int_{E}^{\infty} \! \tilde{I}_{\vec{x}}(\lambda_{\mathrm {D}}(E,E_{S})) \frac{\mathrm{d}n}{\mathrm{d}E}(E_{S}) \, \mathrm{d}E_{S},
\end{equation}
where $ \kappa = (1/2) \left\langle \sigma v \right\rangle (\rho_{\odot}/m_{\mathrm{DM}})^{2} $, with $\left\langle \sigma v \right\rangle$ the annihilation cross section, $\rho_{\odot}$ the DM density in the Solar neighborhood, and $m_{\mathrm{DM}}$ the DM mass. $ \tilde{I}_{\vec{x}} $ is the so-called halo function that contains all the information on diffusion, through the diffusion length $ \lambda_{\mathrm {D}}(E,E_{\mathrm{S}}) $ that depends on the injection energy $E_{\mathrm{S}}$ and the energy after propagation $E$. The halo function is convolved with the injection spectrum $\mathrm{d}n/\mathrm{d}E$. The most difficult step of the resolution is to compute the halo function in the context of a cuspy DM profile. For that we used the method relying on Green's functions. The halo function is thus given by the convolution over the diffusion zone (DZ) of the Green's function $G(\vec{x},E ; \vec{x}_{\mathrm{S}},E_{\mathrm{S}}) \equiv G(\vec{x},\vec{x}_{\mathrm{S}},\lambda_{\mathrm {D}}(E,E_{S}))$ of the diffusion-loss equation and the square of the DM density \cite{Itilde_Timur}
\begin{equation}
\tilde{I}_{\vec{x}}(\lambda_{\mathrm {D}}(E,E_{S})) = \int_{\mathrm{DZ}} \! \mathrm{d}\vec{x}_{\mathrm{S}} \, G(\vec{x},E ; \vec{x}_{\mathrm{S}},E_{\mathrm{S}}) \left( \frac{\rho(\vec{x}_{\mathrm{S}})}{\rho_{\odot}} \right) ^{2}.
\end{equation}
The difficulty with this integral is twofold. First of all, to deal with the steepness of the DM profile, we used logarithmic steps. Second, the halo function must boil down to $(\rho/\rho_{\odot})^{2}$ when diffusion becomes negligible, that is when $\lambda_{\mathrm {D}} \rightarrow 0$ (i.e.~$E \rightarrow E_{\mathrm{S}}$). The problem is that in this regime, $ G $ becomes infinitely peaked, and the sampling of the integral must be carefully chosen in order not to miss the peak. The solution is to define different regimes for $G$ depending on the ratio of $\lambda_{\mathrm {D}}$ and the distance to the GC, as described in detail in \cite{spike_my_paper}. Once $\tilde{I}$ has been computed with this dedicated treatment of diffusion on small scales, one can compute the IC and bremsstrahlung gamma-ray fluxes from electrons and positrons.

\section{Fitting the Fermi-LAT GeV excess with the leptonic channels}

\begin{figure}[!ht]
\begin{center}
\includegraphics[scale=0.37]{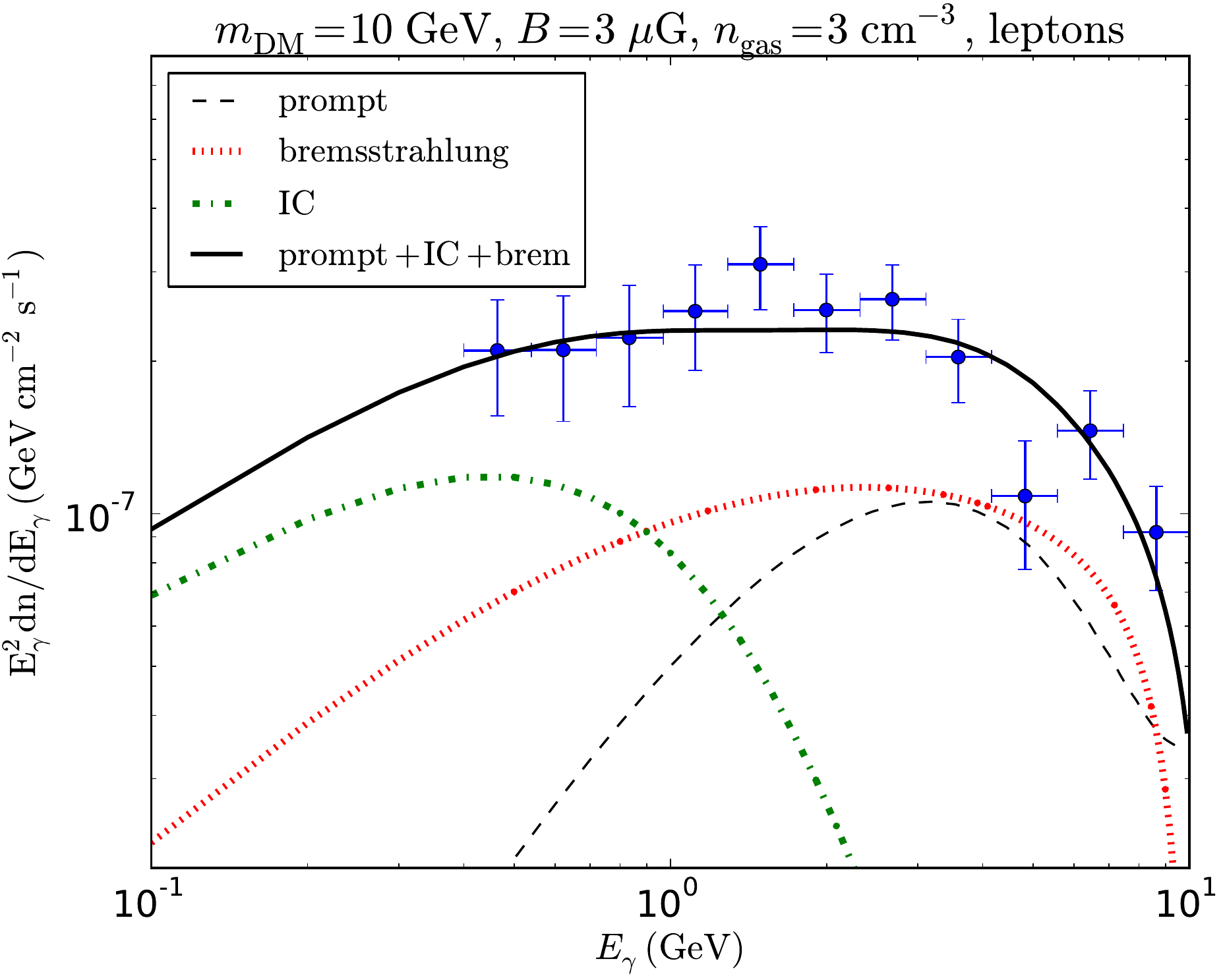}
\caption{\label{leptons_3contributions}Spectrum of the residual extended emission in the $7^{\circ} \times 7^{\circ}$ region around the GC. The data points are taken from \cite{Gordon_Macias_excess}. The prompt (black dashed), IC (green dashed-dotted) and bremsstrahlung (red dotted) emissions from $10\ \rm GeV$ DM democratically annihilating into leptons add up to give a very good fit to the data, as shown by the black solid line, with a best-fit cross section of $\left\langle \sigma v \right\rangle = 0.86 \times 10^{-26}\ \rm cm^{3}\ s^{-1}$.}
\end{center}
\end{figure}

\begin{figure}[!ht]
\centering
\includegraphics[scale=0.37]{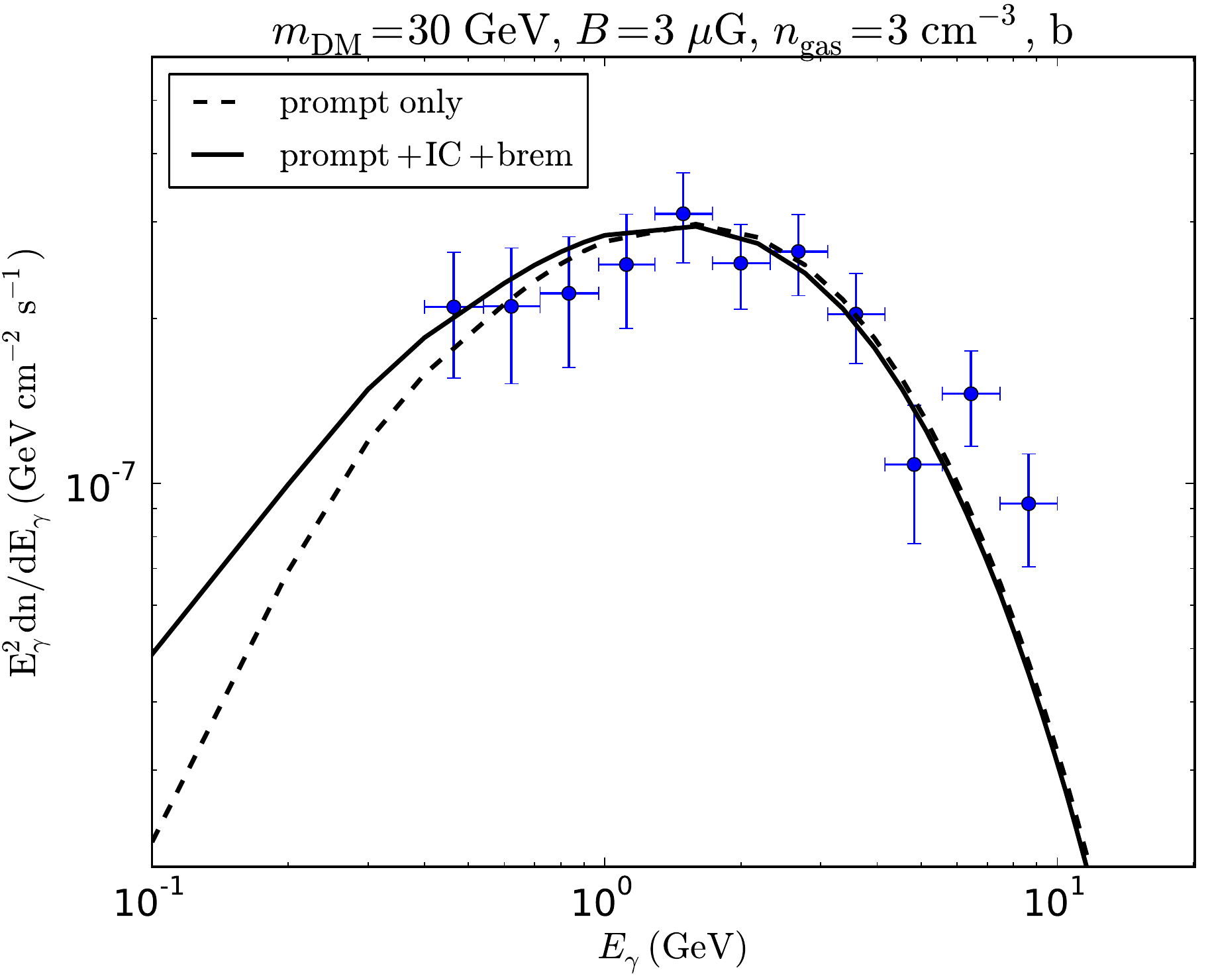}
\includegraphics[scale=0.37]{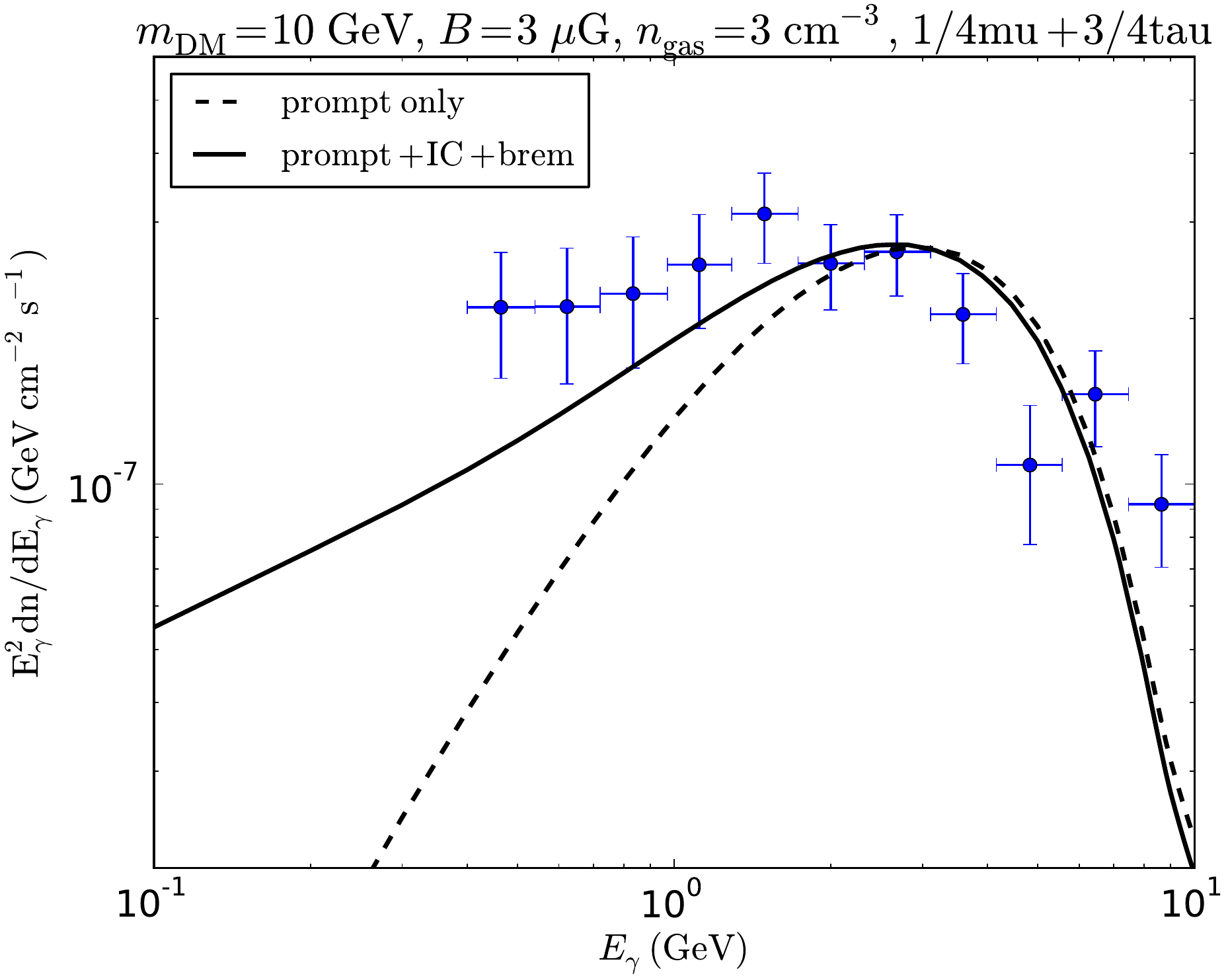}
\caption{\label{best_fits}Best fits to the Fermi residual with the gamma-ray spectrum from annihilations of $30\ \rm GeV$ DM particles into 100\% $b\bar{b}$ (left panel), and 10 GeV DM annihilating into 25\% $\mu^{+}\mu^{-}$ and 75\% $\tau^{+}\tau^{-}$ (right panel). The best-fit cross section is $\sim 2 \times 10^{-26}\ \rm cm^{3}\ s^{-1}$ for the left panel, and $\sim 1 \times 10^{-26}\ \rm cm^{3}\ s^{-1}$ for the right panel.}
\end{figure}

We used fixed values for the quantities describing the interstellar medium (the magnetic field and the gas density) that correspond to average values for the losses in the GC region. In Fig.~\ref{leptons_3contributions} we show that for democratic annihilation into leptons, the contributions from prompt emission, IC and bremsstrahlung are of the same order of magnitude, and they combine to give an excellent fit to the excess, with a best-fit cross section of $\left\langle \sigma v \right\rangle = 0.86 \times 10^{-26}\ \rm cm^{3}\ s^{-1}$. This is a very important result, since it means that 30 GeV DM annihilating into $b\bar{b}$ is not the only possible annihilation set-up, and that DM can annihilate into leptons. This result is discussed in more detail in Ref.~\cite{GeV_excess_my_paper}. We also show in Fig.~\ref{best_fits} (left panel) that including the contributions from IC and bremsstrahlung does not significantly affect the spectrum for the $b\bar{b}$ channel, except at low energies.

Therefore, the effect of including the IC and bremsstrahlung contributions is maximal for democratic annihilation into leptons. However, the authors of Ref.~\cite{Bringmann_constraints} used the AMS data to set constraints on the leptonic channels, which exclude annihilations into $e^{+}e^{-}$ and impose that the branching ratio into $\mu^{+}\mu^{-}$ should not exceed 25\%. We do not discuss the validity of these constraints here, but we show in Fig.~\ref{best_fits} (right panel) the best fit to the excess obtained with a final state of DM annihilation containing 25\% $\mu^{+}\mu^{-}$ and 75\% $\tau^{+}\tau^{-}$. With such branching ratios, the fit is marginally good, with an effect of the secondary contributions to the gamma-ray spectrum less significant than for democratic annihilation.

\begin{figure*}[!ht]
\centering
\includegraphics[scale=0.37]{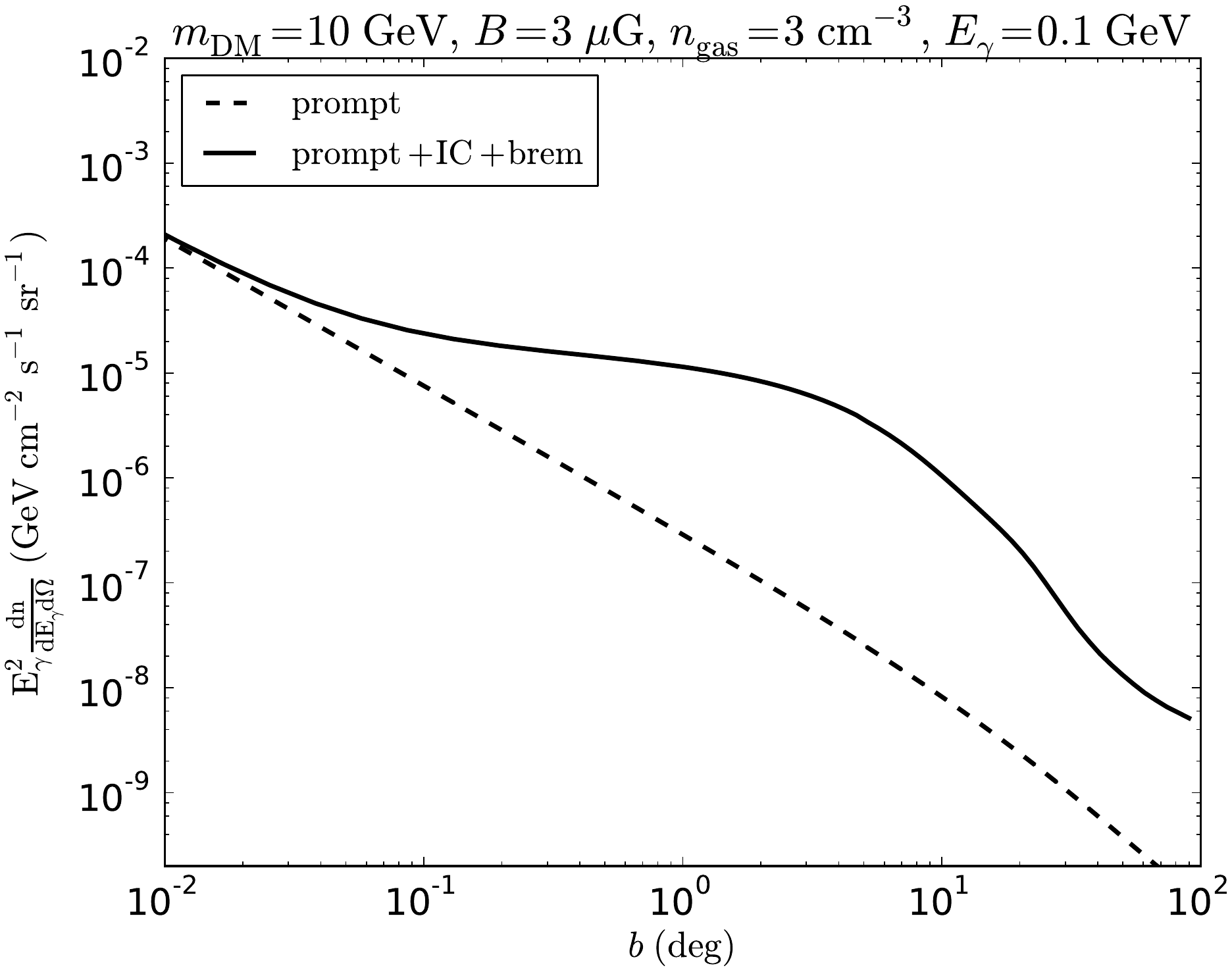}
\includegraphics[scale=0.37]{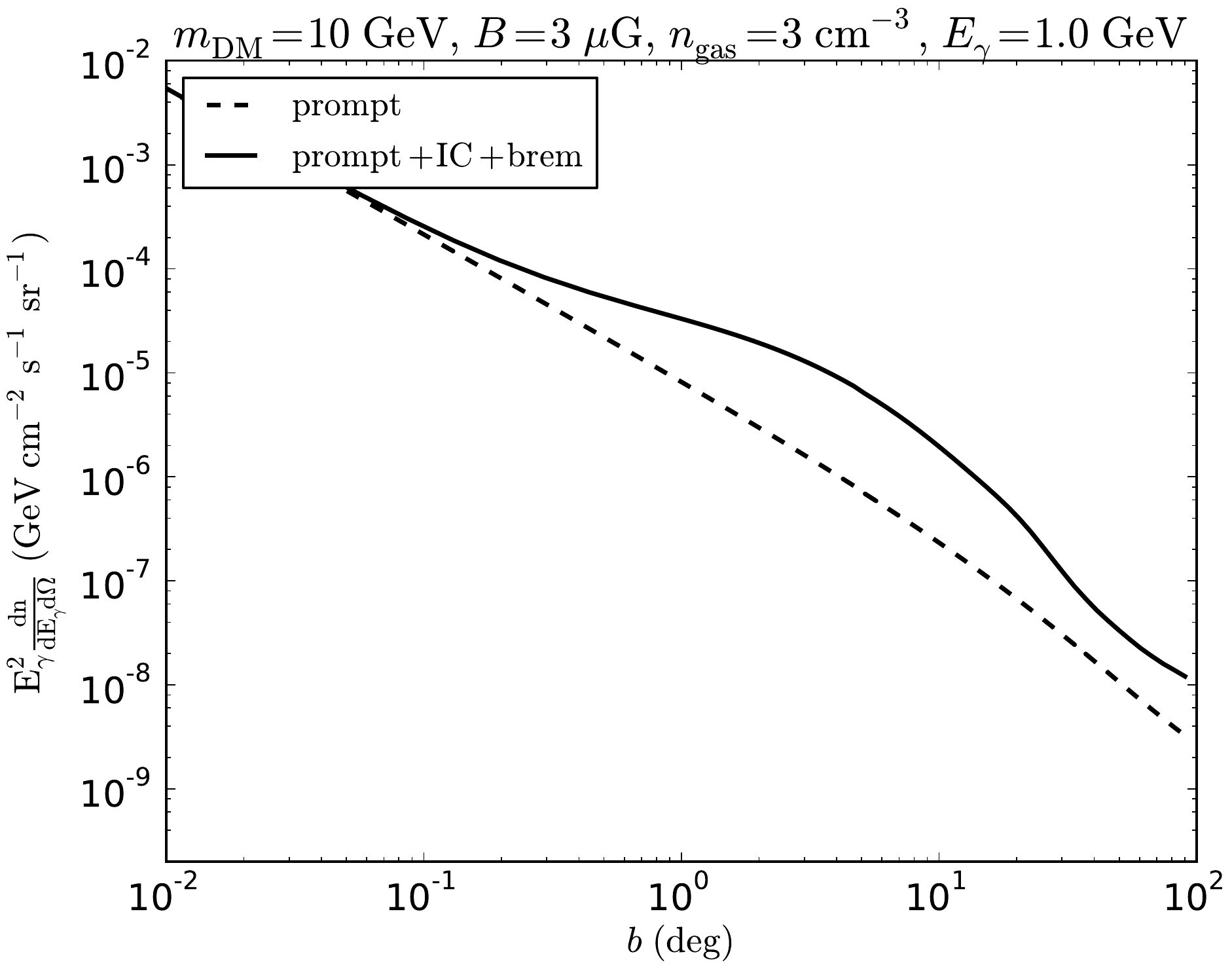}
\caption{\label{flux_vs_b}Gamma-ray flux from DM annihilating exclusively into leptons democratically, as a function of latitude $b$, for gamma-ray energies of 0.1 GeV (left panel) and 1 GeV (right panel).}
\end{figure*}

Finally, a very important test of the leptonic scenario is the morphology of the emission. Indeed, due to spatial diffusion, the IC and bremsstrahlung emissions are more extended than the prompt component, and their morphology depends on the observed energy. This may allow one to set constraints on the annihilation channel. Shown in Fig.~\ref{flux_vs_b} (left panel) is the gamma-ray flux plotted against latitude, for the best-fit parameters corresponding to the spectrum of Fig.~\ref{leptons_3contributions}, and $E_{\gamma} = 0.1\ \rm GeV$. Above a few degrees the morphology should be compatible with the one found in the literature, and corresponding to prompt emission. However, between ${\cal{O}}(0.1)$ and ${\cal{O}}(1)^{\circ}$, the flux profile is shallower than that of the prompt component, which might lead to a tension between the flux from the leptonic channels and the morphology from the literature. However, 0.1 GeV is below the lowest energy data point (around 0.3--0.4 GeV), and the secondary emissions dominate at low energy. It turns out that at 1 GeV, the tension between the total flux from the leptonic channels and the morphology of the prompt emission is much weaker, as shown in Fig.~\ref{flux_vs_b} (right panel).

\section*{Conclusion}

The Fermi-LAT GeV excess is a strong case for DM, and annihilations of 30 GeV particles into $b\bar{b}$ provide the simplest set-up a priori. However, it is very important to take into account all relevant emission processes and spatial diffusion. Including all these processes drastically changes the interpretation of the excess in terms of DM. Therefore, we showed that $b\bar{b}$ is not the only possible channel and DM can in fact annihilate into leptons. Finally, the morphology below $\sim 1^{\circ}$ at low energies can help to discriminate between the leptonic and $b\bar{b}$ channels.

\begin{small}
\acknowledgments
I thank C\'{e}line B{\oe}hm and Joseph Silk for a fruitful collaboration on this project. This research has been supported at IAP by the ERC Project No.~267117 (DARK) hosted by Universit\'e Pierre et Marie Curie (UPMC) - Paris 6 and at JHU by NSF Grant No.~OIA-1124403. This work has been also supported by UPMC and STFC. Finally, this project has been carried out in the ILP LABEX (ANR-10-LABX-63) and has been supported by French state funds managed by the ANR, within the Investissements d'Avenir programme (ANR-11-IDEX-0004-02).
\end{small}


\begin{thebibliography}{99}
  \bibitem{Fermi_collaboration} V. Vitale and A. Morselli,
  \emph{Indirect Search for Dark Matter from the center of the Milky Way with the Fermi-Large Area Telescope},
  \emph{ArXiv e-prints} (2009)
  [{\tt astro-ph/0912.3828}].
  
  \bibitem{Hooper_Linden_excess} D. Hooper and T. Linden,
  \emph{On the Origin of the Gamma Rays from the Galactic Center},
  \emph{Phys. Rev. D} {\bf 84} (2011) 123005
  [{\tt astro-ph/1110.0006}].
  
  \bibitem{Gordon_Macias_excess} C. Gordon and O. Macias,
  \emph{Dark Matter and Pulsar Model Constraints from Galactic Center Fermi-LAT Gamma Ray Observations},
  \emph{Phys. Rev. D} {\bf 88} (2013) 083521
  [{\tt astro-ph/1306.5725}].
  
  \bibitem{Abazajian_GeV_excess} K.~N. Abazajian, N. Canac, S. Horiuchi and M. Kaplinghat,
  \emph{Astrophysical and Dark Matter Interpretations of Extended Gamma Ray Emission from the Galactic Center},
  \emph{Phys. Rev. D} {\bf 90} (2014) 023526
  [{\tt astro-ph/1402.4090}].
  
  \bibitem{Daylan_GeV_excess} T. Daylan, D.~P. Finkbeiner, D. Hooper, T. Linden, S.~K.~N. Portillo, N.~L. Rodd and T.~R. Slatyer,
  \emph{The Characterization of the Gamma-Ray Signal from the Central Milky Way: A Compelling Case for Annihilating Dark Matter},
  \emph{ArXiv e-prints} (2014)
  [{\tt astro-ph/1402.6703}].
  
  \bibitem{Fermi_IC} Fermi-LAT Collaboration,
  \emph{Constraints on the Galactic Halo Dark Matter from Fermi-LAT Diffuse Measurements},
  \emph{ApJ} {\bf 761} (2012) 91
  [{\tt astro-ph/1205.6474}].
	
  \bibitem{brems_Cirelli} M. Cirelli, P.~D. Serpico and G. Zaharijas,
  \emph{Bremsstrahlung gamma rays from light dark matter},
  \emph{JCAP} {\bf 11} (2013) 35
  [{\tt astro-ph/1307.7152}].
  
  
  \bibitem{Itilde_Timur} T. Delahaye, R. Lineros, F. Donato, N. Fornengo and P. Salati,
  \emph{Positrons from Dark Matter Annihilation in the Galactic Halo: Theoretical Uncertainties},
  \emph{Phys. Rev. D} {\bf 77} (2008) 063527
  [{\tt astro-ph/0712.2312}].
  
  \bibitem{spike_my_paper} T. Lacroix, C. B{\oe}hm and J. Silk,
  \emph{Probing a dark matter density spike at the Galactic Center},
  \emph{Phys. Rev. D} {\bf 89} (2014) 063534
  [{\tt astro-ph/1311.0139}].

  \bibitem{GeV_excess_my_paper} T. Lacroix, C. B{\oe}hm and J. Silk,
  \emph{Fitting the Fermi-LAT GeV excess: On the importance of including the propagation of electrons from dark matter},
  \emph{Phys. Rev. D} {\bf 90} (2014) 043508
  [{\tt astro-ph/1403.1987}].

  \bibitem{Bringmann_constraints} T. Bringmann, M. Vollmann and C. Weniger,
  \emph{Updated cosmic-ray and radio constraints on light dark matter: Implications for the GeV gamma-ray excess at the Galactic     
  center},
  \emph{ArXiv e-prints} (2014)
  [{\tt astro-ph/1406.6027}].

  
\end{thebibliography}
\end{document}